\documentclass[11pt]{article}
\usepackage{amsfonts,amsmath,amssymb}
\usepackage{algorithm2e,framed}

\newcommand{\Probab}[1]{\mbox{}{\bf{Pr}}\left[#1\right]}
\newcommand{\Expect}[1]{\mbox{}{\bf{E}}\left[#1\right]}
\newcommand{\Varnce}[1]{\mbox{}{\bf{Var}}\left[#1\right]}

\newcommand{\FNorm }[1]{\mbox{}\left\|#1\right\|_F  }
\newcommand{\FNormS}[1]{\mbox{}\left\|#1\right\|_F^2}

\newcommand{\TNorm }[1]{\mbox{}\left\|#1\right\|_2  }
\newcommand{\TNormS}[1]{\mbox{}\left\|#1\right\|_2^2}

\newcommand{\VTTNorm }[1]{\mbox{}\left\|#1\right\|_2  }
\newcommand{\VTTNormS}[1]{\mbox{}\left\|#1\right\|_2^2}

\newcommand{\VINorm }[1]{\mbox{}\left\|#1\right\|_{\infty}  }

\newcommand{\setlinespacing}[1]%
           {\setlength{\baselineskip}{#1 \defbaselineskip}}

\newcommand{\abs }[1]{\left|#1\right|}

\newtheorem{lemma}{Lemma}
\newtheorem{theorem}{Theorem}

\newenvironment{Proof}{\noindent {\em Proof:}}{\\\hspace*{\fill}\mbox{$\diamond$}}

\long\def\killtext#1{}

\def\Prob{\hbox{\bf{Pr}}}

\newlength{\defbaselineskip}
\begin{document}

\title{Faster Least Squares Approximation}

\author{
Petros Drineas
\thanks{
Department of Computer Science,
Rensselaer Polytechnic Institute,
Troy, NY,
drinep@cs.rpi.edu.
}
\and Michael W. Mahoney
\thanks{
Department of Mathematics,
Stanford University,
Stanford, CA,
mmahoney@cs.stanford.edu.
}
\and S. Muthukrishnan
\thanks{
Google, Inc., New York, NY, muthu@google.com.}
\and Tam\'{a}s Sarl\'{o}s
\thanks{
Yahoo! Research, Sunnyvale, CA, stamas@yahoo-inc.com.}
}

\date{}
\maketitle
\vspace{5mm}

\begin{abstract}
Least squares approximation is a technique to find an approximate solution to a system of linear equations that has no exact solution. In a typical setting, one lets $n$ be the number of constraints and $d$ be the number of variables, with $n \gg d$. Then, existing exact methods find a solution vector in $O(nd^2)$ time. We present two randomized algorithms that provide accurate relative-error approximations to the optimal value and the solution vector of a least squares approximation problem more rapidly than existing exact algorithms. Both of our algorithms preprocess the data with the Randomized Hadamard Transform. One then uniformly randomly samples constraints and solves the smaller problem on those constraints, and the other performs a sparse random projection and solves the smaller problem on those projected coordinates. In both cases, solving the smaller problem provides relative-error approximations, and, if $n$ is sufficiently larger than $d$, the approximate solution can be computed in $O(nd \ln d)$ time.
\end{abstract}

\vspace{5mm}
\section{Introduction}
\label{sxn:intro}

In many applications in mathematics and statistical data analysis, it is of interest to find an approximate solution to a system of linear equations that has no exact solution. For example, let a
matrix $A \in \mathbb{R}^{n \times d}$ and a vector $b \in \mathbb{R}^n$ be given. If $n \gg d$, there will not in general exist a vector $x \in \mathbb{R}^d$ such that $Ax=b$, and yet it is
often of interest to find a vector $x$ such that $Ax \approx b$ in some precise sense. The method of least squares, whose original formulation is often credited to Gauss and Legendre~\cite{Stigler86}, accomplishes this by minimizing the sum of squares of the elements of the residual vector, i.e., by solving the optimization problem
\begin{equation}
\label{eqn:orig_ls_prob}
\mathcal{Z} = \min_{x \in \mathbb{R}^d}
\VTTNorm{Ax - b}.
\end{equation}
It is well-known that the minimum $\ell_2$-norm vector among those satisfying eqn.~(\ref{eqn:orig_ls_prob}) is
\begin{equation}
\label{eqn:xopt_orig_ls_prob}
x_{opt} = A^{\dagger}b,
\end{equation}
where $A^{\dagger}$ denotes the Moore-Penrose generalized inverse of the matrix $A$~\cite{BIG03,GVL96}. This solution vector has a very natural statistical interpretation as providing an optimal estimator among all linear unbiased estimators, and it has a very natural geometric interpretation as providing an orthogonal projection of the vector $b$ onto the span of the columns of the matrix $A$.

Recall that to minimize the quantity in eqn.~(\ref{eqn:orig_ls_prob}), we can set the derivative of $\VTTNormS{Ax-b}=(Ax-b)^T(Ax-b)$ with respect to $x$ equal to zero, from which it follows that the minimizing vector $x_{opt}$ is a solution of the so-called normal equations
\begin{equation}
\label{eqn:normal_eqn}
A^TAx_{opt}=A^Tb  .
\end{equation}
Geometrically, this means that the residual vector $b^{\perp}=b-Ax_{opt}$ is required to be orthogonal to the column space of $A$, i.e., ${b^{\perp}}^TA=0$. While solving the normal equations squares the condition number of the input matrix (and thus is not recommended in practice), direct methods (such as the QR decomposition~\cite{GVL96}) solve the problem of eqn.~(\ref{eqn:orig_ls_prob}) in $O(nd^2)$ time assuming that $n \geq d$. Finally, an alternative expression for the vector $x_{opt}$ of eqn.~(\ref{eqn:xopt_orig_ls_prob}) emerges by leveraging the Singular Value Decomposition (SVD) of $A$. If $A = U_A\Sigma_A V_A^T$ denotes the SVD of $A$, then
\begin{equation*}
x_{opt}=V_A\Sigma_A^{-1}U_A^Tb.
\end{equation*}

\subsection{Our results}\label{sxn:intro:ourresults}

In this paper, we describe two randomized algorithms that will provide accurate relative-error approximations to the minimal $\ell_2$-norm solution vector $x_{opt}$ of eqn.~(\ref{eqn:xopt_orig_ls_prob}) faster than existing exact algorithms for a large class of overconstrained least-squares problems. In particular, we will prove the following theorem.

\begin{theorem}
\label{thm:main_result}
Suppose $A \in \mathbb{R}^{n \times d}$, $b \in \mathbb{R}^{n}$, and let $\epsilon \in (0,1)$. Then, there exists a randomized algorithm that returns a vector $\tilde{x}_{opt} \in \mathbb{R}^d$ such that, with probability at least $.8$, the following two claims hold: first, $\tilde{x}_{opt}$ satisfies
\begin{equation}
\label{eqn:result1_intro}
\VTTNorm{A\tilde{x}_{opt}-b} \le (1+\epsilon) \mathcal{Z}  ;
\end{equation}
and, second, if $\kappa(A)$ is the condition number of $A$ and if we assume that $\gamma \in [0,1]$ is the fraction of the norm of $b$ that lies in the column space of $A$ (i.e., $\gamma = \VTTNorm{U_A U_A^T b}/\VTTNorm{b}$, where $U_A$ is an orthogonal basis for the column space of $A$), then $\tilde{x}_{opt}$ satisfies
\begin{equation}
\label{eqn:result2_intro}
\VTTNorm{x_{opt}-\tilde{x}_{opt}}
  \leq \sqrt{\epsilon}
       \left(\kappa(A)
       \sqrt{\gamma^{-2}-1}\right)
       \VTTNorm{x_{opt}}.
\end{equation}
Finally, the solution $\tilde{x}_{opt}$ can be computed in $O(nd \ln d)$ time if $n$ is sufficiently larger than $d$ and less than $e^d$.
\end{theorem}
We will provide a precise statement of the running time for our two algorithms (including the $\epsilon$-dependence) in Theorems~\ref{thm:alg_sample_fast} (Section~\ref{sxn:sampling})
and~\ref{thm:alg_projection_fast} (Section~\ref{sxn:projection}), respectively. It is worth noting that the claims of Theorem~\ref{thm:main_result} can be made to hold with probability $1-\delta$, for any $\delta>0$, by repeating the algorithm $\left\lceil \ln(1/\delta)/\ln(5)\right\rceil$ times. For example, one could run ten independent copies of the algorithm and keep the vector $\tilde{x}_{opt}$ that minimizes the residual. This clearly does not increase the running time of the algorithm by more than a constant factor, while driving the failure probability down to (approximately) $10^{-7}$. Also, we will assume that $n$ is a power of two and that the rank of the $n \times d$ matrix $A$ equals $d$. (We note that padding $A$ and $b$ with all-zero rows suffices to remove the first assumption.)

We now provide a brief overview of our main algorithms. Let the matrix product $HD$ denote the $n \times n$ Randomized Hadamard Transform (see also Section~\ref{sxn:RHT}). Here the $n \times n$ matrix $H$ denotes the (normalized) matrix of the Hadamard transform and the $n \times n$ diagonal matrix $D$ is formed by setting its diagonal entries to $+1$ or $-1$ with equal probability in $n$  independent trials. This transform has been used as one step in the development of a ``fast'' version of the Johnson-Lindenstrauss lemma~\cite{AC06,Matousek08_RSA}. Our first algorithm is a random sampling algorithm. After premultiplying $A$ and $b$ by $HD$, this algorithm samples uniformly at random $r$ constraints from the preprocessed problem. (See eqn.~(\ref{eqn:rvaluefinal}), as well as the remarks after Theorem~\ref{thm:alg_sample_fast} for the precise value of $r$.) Then, this algorithm solves the least squares problem on just those sampled constraints to obtain a vector $\tilde{x}_{opt} \in \mathbb{R}^d$ such that Theorem~\ref{thm:main_result} is satisfied. Note that applying the randomized Hadamard transform to the matrix $A$ and vector $b$ only takes $O(n d \ln r)$ time. This follows since we will actually sample only $r$ of the constraints from the Hadamard-preprocessed problem~\cite{AL08}. Then, exactly solving the $r \times d$ sampled least-squares problem will require only $O(rd^2)$ time. Assuming that $\epsilon$ is a constant and $n \leq e^d$, it follows that the running time of this algorithm is $O(nd\ln d)$ when $\frac{n}{\ln n} = \Omega(d^2)$.

In a similar manner, our second algorithm also initially premultiplies $A$ and $b$ by $HD$. This algorithm then multiplies the result by a $k \times n$ sparse projection matrix $T$, where $k=O(d/\epsilon)$. This matrix $T$ is described in detail in Section~\ref{sxn:review_previous:projecting}. Its construction depends on a sparsity parameter, and it is identical to the ``sparse
projection'' matrix in Matou\v{s}ek's version of the Ailon-Chazelle result~\cite{AC06,Matousek08_RSA}. Finally, our second algorithm solves the least squares problem on just those $k$ coordinates to
obtain $\tilde{x}_{opt} \in \mathbb{R}^d$ such that the three claims of Theorem~\ref{thm:main_result} are satisfied. Assuming that $\epsilon$ is a constant and $n \leq e^d$, it follows that the running time of this algorithm is $O(nd\ln d)$ when $n = \Omega(d^2)$.

It is worth noting that our second algorithm has a (marginally) less restrictive assumption on the connection between $n$ and $d$. However, the first algorithm is simpler to implement and easier to describe. Clearly, an interesting open problem is to relax the above constraints on $n$ for either of the proposed algorithms.

\subsection{Related work}

We should note several lines of related work.
\begin{itemize}
\item
First, techniques such as the ``method of averages''~\cite{DSS68} preprocess
the input into the form of eqn.~(\ref{eqn:orig_ls_prob_Xrotated}) of
Section~\ref{sxn:precond} and can be used to
obtain exact or approximate solutions to the least squares problem of
eqn.~(\ref{eqn:orig_ls_prob}) in $o(nd^2)$ time under strong statistical
assumptions on $A$ and $b$.
To the best of our knowledge, however, the two algorithms we present and
analyze are the first algorithms to provide nontrivial approximation
guarantees for overconstrained least squares approximation problems in
$o(nd^2)$ time, while making no assumptions at all on the input data.
\item
Second, Ibarra, Moran, and Hui~\cite{IMH82} provide a reduction of the least
squares approximation problem to the matrix multiplication problem.
In particular, they show that $MM(d) O(n/d)$ time, where $MM(d)$ is the time
needed to multiply two $d \times d$ matrices, is sufficient to solve this
problem.
All of the running times we report in this paper assume the use of standard
matrix multiplication algorithms, since $o(d^3)$ matrix multiplication
algorithms are almost never used in practice.
Moreover, even with the current best value for the matrix multiplication
exponent, $\omega \approx 2.376$~\cite{CW87}, our algorithms are still
faster.
\item
Third, motivated by our preliminary results as reported in~\cite{DMM06}
and~\cite{Sarlos06}, both Rokhlin and Tygert~\cite{RT08} as well as Avron,
Maymounkov, and Toledo~\cite{AMT09_DRAFT,AMT10} have empirically evaluated
numerical implementations of variants of one of the algorithms we introduce.
We describe this in more detail below in Section~\ref{sxn:intro-empirical}.
\item
Fourth, very recently, Clarkson and Woodruff proved space lower bounds on
related problems~\cite{CW09}; and Nguyen, Do, and Tran achieved a small
improvement in the sampling complexity for related
problems~\cite{NDT09}.
\end{itemize}

\subsection{Empirical performance of our randomized algorithms}
\label{sxn:intro-empirical}

In prior work we have empirically evaluated randomized algorithms that rely on
the ideas that we introduce in this paper in several large-scale data
analysis tasks. Nevertheless, it is a fair question to ask whether our ``random
perspective'' on linear algebra will work well in numerical
implementations of interest in scientific computation.
We address this question here.
Although we do \emph{not} provide an empirical evaluation in this paper,
in the wake of the original Technical Report version of this paper in
2007~\cite{DMMS07_FastL2_TR}, two groups of researchers have
demonstrated that numerical implementations of variants of the
algorithms we introduce in this paper can perform very well in practice.
\begin{itemize}
\item
In 2008, Rokhlin and Tygert~\cite{RT08} describe a variant of our random
projection algorithm, and they demonstrate that their algorithm runs in
time
$$
O(\ln(\ell) + \kappa\ln(1/\epsilon)nd+d^2\ell) ,
$$
where $\ell$ is an ``oversampling'' parameter and $\kappa$ is a condition
number.
Importantly (at least for very high-precision applications of this random
sampling methodology), they reduce the dependence on $\epsilon$ from
$1/\epsilon$ to $\ln(1/\epsilon)$.
Moreover, by choosing $\ell \ge 4d^2$, they demonstrate that $\kappa \le 3$.
Although this bound is inferior to ours, they also consider a class of
matrices for which choosing $\ell=4d$ empirically produced a condition
number $\kappa <3$, which means that for this class of matrices their
running time is
$$
O(\ln(d) + \kappa\ln(1/\epsilon)nd+d^3) .
$$
Their numerical experiments on this class of matrices clearly indicate that
their implementations of variants of our algorithms
perform well for certain matrices as small as thousands of rows by hundreds
of columns.
\item
In 2009, Avron, Maymounkov, Toledo~\cite{AMT09_DRAFT,AMT10}
introduced a randomized least-squares solver based directly on our
algorithms.
They call it Blendenpik, and by considering a much broader class of
matrices, they demonstrate that their solver ``beats \textsc{LAPACK}'s
direct dense least-sqares solver by a large margin on essentially any
dense tall matrix.''
Beyond providing additional theoretical analysis, including backward error
analysis bounds for our algorithm,
they consider five (and numerically implement three) random projection
strategies (i.e., Discrete Fourier Transform, Discrete Cosine
Transform, Discrete Hartely Transform, Walsh-Hadamard Transform, and a Kac
random walk), and they evaluate their algorithms on a wide range of matrices
of various sizes and various ``localization '' or ``coherence'' properties.
Based on these results that empirically show the superior performance of
randomized algorithms such as those we introduce and analyze in this paper
on a wide class of matrices, they go so far as to ``suggest that
random-projection algorithms should be incorporated into future versions of
\textsc{LAPACK}.''
\end{itemize}

\subsection{Outline}

After a brief review of relevant background in Section~\ref{sxn:review_la}, Section~\ref{sxn:precond} presents a structural result outlining conditions on preconditioner matrices that are sufficient for relative-error approximation. Then, we present our main sampling-based algorithm for approximating least squares approximation in Section~\ref{sxn:sampling} and in Section~\ref{sxn:projection} we present a second projection-based algorithm for the same problem.
Preliminary versions of parts of this paper have appeared as conference
proceedings in the 17th ACM-SIAM Symposium on Discrete
Algorithms~\cite{DMM06} and in the 47th IEEE Symposium on Foundations of
Computer Science~\cite{Sarlos06}; and the original Technical Report version
of this journal paper has appeared on the arXiv~\cite{DMMS07_FastL2_TR}.
In particular, the core of our analysis in this paper was introduced in~\cite{DMM06}, where an expensive-to-compute probability distribution was used to construct a relative-error approximation sampling algorithm for the least squares approximation problem. Then, after the development of the Fast Johnson-Lindenstrauss transform~\cite{AC06}, \cite{Sarlos06} proved that similar ideas could be used to improve the running time of randomized algorithms for the least squares approximation problem. In this paper, we have combined these ideas, treated the two algorithms in a manner to highlight their similarities and differences, and considerably simplified the analysis.

\section{Preliminaries}
\label{sxn:review_la}

\subsection{Notation}

We let $[n]$ denote the set $\{1,2,\ldots,n\}$; $\ln x$ denotes the natural logarithm of $x$ and $\log_2 x$ denotes the base two logarithm of $x$. For any matrix $A \in \mathbb{R}^{n \times d}$, $A_{(i)}, i \in [n]$ denotes the $i$-th row of $A$ as a row vector and $A^{(j)}, j \in [d]$ denotes the $j$-th column of $A$ as a column vector. Also, given a random variable $X$, we let $\Expect{X}$ denote its expectation and $\Varnce{X}$ denote its variance.

We will make frequent use of matrix and vector norms. More specifically, we let $$ \FNormS{A} = \sum_{i=1}^n \sum_{j=1}^d A_{ij}^2 $$ denote the square of the Frobenius norm of $A$, and we let $$\TNorm{A} = \sup_{x\in \mathbb{R}^d,\ \VTTNorm{x}=1} \VTTNorm{Ax}$$ denote the spectral norm of $A$. For any vector $x \in \mathbb{R}^n$, its $\ell_2$-norm (or Euclidean norm) is equal to the square root of the sum of the squares of its elements, while its $\ell_\infty$ norm is defined as $\VINorm{x} = \max_{i \in [n]} \abs{x_i}$.

\subsection{Linear Algebra background}

We now review relevant definitions and facts from linear algebra; for more details, see~\cite{Stewart90,GVL96,Bhatia97,BIG03}. Let the rank of $A \in \mathbb{R}^{n \times d}$ be $\rho \leq \min\{n,d\}$. The Singular Value Decomposition (SVD) of $A$ is denoted by $A = U_A \Sigma_A V_A^T$, where $U_A \in \mathbb{R}^{n \times \rho}$ is the matrix of left singular vectors, $\Sigma_A \in \mathbb{R}^{\rho \times \rho}$ is the diagonal matrix of non-zero singular values, and $V_A \in \mathbb{R}^{d \times \rho}$ is the matrix of right singular vectors. Let $\sigma_i(A), i\in [\rho]$, denote the $i$-th non-zero singular value of $A$, and $\sigma_{\max}(A)$ and $\sigma_{\min}(A)$ denote the maximum and minimum singular value of $A$. The condition number of $A$ is $\kappa(A) = \sigma_{\max}(A)/\sigma_{\min}(A)$. The Moore-Penrose generalized inverse, or pseudoinverse, of $A$ may be expressed in terms of the SVD as $A^{\dagger}=V_A\Sigma_A^{-1}U_A^T$~\cite{BIG03}. Finally, for any orthogonal matrix $U \in \mathbb{R}^{n \times \ell}$, let $U^{\perp} \in \mathbb{R}^{n \times (n-\ell)}$ denote an orthogonal matrix whose columns are an orthonormal basis spanning the subspace of $\mathbb{R}^n$ that is orthogonal to the column space of $U$.  In terms of $U_A^{\perp}$, the optimal value of the least squares residual of eqn.~(\ref{eqn:orig_ls_prob}) is
$$
\mathcal{Z} = \min_{x \in \mathbb{R}^d} \VTTNorm{Ax-b}
            = \VTTNorm{U_A^{\perp}{U_A^{\perp}}^Tb}         .
$$
\subsection{Markov's inequality and the union bound}

We will make frequent use of the following fundamental result from probability theory, known as Markov's inequality~\cite{MotwaniRaghavan95}. Let $X$ be a random variable assuming non-negative values with expectation $\Expect{X}$. Then, for all $t > 0$,
$$X \leq t \cdot \Expect{X}$$
with probability at least $1-t^{-1}$.

We will also need the so-called union bound. Given a set of random events ${\cal E}_1,{\cal E}_2,\ldots,{\cal E}_{n}$ holding with respective probabilities $p_1,p_2,\ldots,p_n$, the probability that all events hold (i.e., the probability of the union of those events) is upper bounded by $\sum_{i=1}^n p_i$.

\subsection{The Randomized Hadamard Transform}\label{sxn:RHT}

The Randomized Hadamard Transform was introduced in~\cite{AC06} as one step in the development of a fast version of the Johnson-Lindenstrauss lemma~\cite{AC06,Matousek08_RSA}. Recall that the (non-normalized) $n \times n$ matrix of the Hadamard transform $H_n$ may be defined recursively as follows:
$$ H_n = \left[
\begin{array}{cc}
  H_{n/2} & H_{n/2} \\
  H_{n/2} & -H_{n/2}
\end{array}\right]   ,
\qquad \mbox{with} \qquad
H_2 = \left[
\begin{array}{cc}
  +1 & +1 \\
  +1 & -1
\end{array}\right].
$$
The $n \times n$ normalized matrix of the Hadamard transform is equal to $\frac{1}{\sqrt{n}}H_n$; hereafter, we will denote this normalized matrix by $H$. Now consider a diagonal matrix $D \in \mathbb{R}^{n \times n}$ such that the diagonal entries $D_{ii}$ are set to +1 with probability $1/2$ and to $-1$ with probability $1/2$ in $n$ independent trials. The product $HD$ is the Randomized Hadamard Transform and has two useful properties. First, when applied to a vector, it ``spreads out'' its energy, in the sense of providing a bound for its infinity norm (see Section~\ref{sxn:review_previous:hadamard}). Second, computing the product $HDx$ for any vector $x \in \mathbb{R}^n$ takes $O(n\log_2 n)$ time. Even better, if we only need to access, say, $r$ elements in the transformed vector, then those $r$ elements can be computed in $O(n \log_2 r )$ time~\cite{AL08}. We will expand on the latter observation in the proofs of Theorems~\ref{thm:alg_sample_fast} and~\ref{thm:alg_projection_fast}.

\section{Our algorithms as preconditioners}\label{sxn:precond}

Both of our algorithms may be viewed as preconditioning the input matrix $A$ and the target vector $b$ with a carefully-constructed data-independent random matrix $X$. For our random sampling algorithm, we let $X = S^THD$, where $S$ is a matrix that represents the sampling operation and $HD$ is the Randomized Hadamard Transform, while for our random projection algorithm, we let $X = THD$, where $T$ is a random projection matrix. Thus, we replace the least squares approximation problem of eqn.~(\ref{eqn:orig_ls_prob}) with the least squares approximation problem
\begin{equation}
\label{eqn:orig_ls_prob_Xrotated}
\tilde{\mathcal{Z}} = \min_{x \in \mathbb{R}^d} \VTTNorm{X (Ax - b)}.
\end{equation}
We explicitly compute the solution to the above problem using a traditional deterministic algorithm~\cite{GVL96}, e.g., by computing the vector
\begin{equation}
\label{eqn:xopt_orig_ls_prob_Xrotated}
\tilde{x}_{opt} = \left(XA\right)^{\dagger}Xb  .
\end{equation}
Alternatively, one could use standard iterative methods such as the the Conjugate Gradient Normal Residual method (CGNR, see~\cite{GVL96} for details), which can produce an $\epsilon$-approximation to the optimal solution of eqn.~(\ref{eqn:orig_ls_prob_Xrotated}) in $O(\kappa(XA) rd \ln(1/\epsilon) )$ time, where $\kappa(XA)$ is the condition number of $XA$ and $r$ is the number of rows of $XA$.

\subsection{A structural result sufficient for relative-error approximation}
\label{sxn:sampling:proofs:structural}

In this subsection, we will state and prove a lemma that establishes sufficient conditions on any matrix $X$ such that the solution vector $\tilde{x}_{opt}$ to the least squares problem of eqn.~(\ref{eqn:orig_ls_prob_Xrotated}) will satisfy relative-error bounds of the form~(\ref{eqn:result1_intro}) and~(\ref{eqn:result2_intro}). Recall that the SVD of $A$ is $A=U_A\Sigma_AV_A^T$. In addition, for notational simplicity, we let $b^{\perp} = U_A^{\perp}{U_A^{\perp}}^{T}b$ denote the  part of the right hand side vector $b$ lying outside of the column space of $A$.

The two conditions that we will require of the matrix $X$ are:
\begin{eqnarray}
\label{eqn:lemma1_ass1}
& & \sigma_{min}^2 \left( XU_A \right) \ge 1/\sqrt{2} \mbox{; and}  \\
\label{eqn:lemma1_ass2} & &
\VTTNormS{U_A^TX^TXb^{\perp}}
      \le \epsilon \mathcal{Z}^2/2  ,
\end{eqnarray}
for some $\epsilon \in (0,1)$. Several things should be noted about these conditions. First, although condition~(\ref{eqn:lemma1_ass2}) depends on the right hand side vector $b$, Algorithms~\ref{alg:alg_sample_fast} and~\ref{alg:alg_projection_fast} will satisfy it without using any information from $b$. Second, although condition~(\ref{eqn:lemma1_ass1}) only states that $\sigma_i^2(XU_A)\geq 1/\sqrt{2}$, for all $i \in [d]$, for both of our randomized algorithms we will show that $\abs{1-\sigma_i^2(XU_A)} \le 1-2^{-1/2}$, for all $i \in [d]$. Thus, one should think of $XU_A$ as an approximate isometry. Third, condition~(\ref{eqn:lemma1_ass2}) simply states that $Xb^{\perp}=XU_A^{\perp}{U_A^{\perp}}^{T}b$ remains approximately orthogonal to $XU_A$. Finally, note that the following lemma is a deterministic statement, since it makes no explicit reference to either of our randomized algorithms. Failure probabilities will enter later when we show that our randomized algorithms satisfy conditions~(\ref{eqn:lemma1_ass1}) and~(\ref{eqn:lemma1_ass2}).

\begin{lemma} \label{lem:suff_cond}
Consider the overconstrained least squares approximation problem of eqn.~(\ref{eqn:orig_ls_prob}) and let the matrix $U_A \in \mathbb{R}^{n \times d}$ contain the top $d$ left singular vectors of $A$. Assume that the matrix $X$ satisfies conditions~(\ref{eqn:lemma1_ass1}) and~(\ref{eqn:lemma1_ass2}) above, for some $\epsilon \in (0,1)$. Then, the solution vector $\tilde{x}_{opt}$ to the least squares approximation
problem~(\ref{eqn:orig_ls_prob_Xrotated}) satisfies:
\begin{eqnarray}
\label{eqn:lemma1_eq3}
\VTTNorm{A\tilde{x}_{opt}-b} &\le& (1+\epsilon) \mathcal{Z}  \mbox{, and}  \\
\label{eqn:lemma1_eq4}
\VTTNorm{x_{opt}-\tilde{x}_{opt}}
  &\leq& \frac{1}{\sigma_{min}(A)}\sqrt{\epsilon}\mathcal{Z}  .
\end{eqnarray}
\end{lemma}
\begin{Proof}
Let us first rewrite the down-scaled regression problem induced by
$X$ as
\begin{eqnarray}
\min_{x \in \mathbb{R}^d} \VTTNormS{ Xb - XAx}
\label{eqn:ds1} &=& \min_{y \in \mathbb{R}^d} \VTTNormS{
X(Ax_{opt}+b^{\perp}) -
XA(x_{opt}+y) }                 \\
\nonumber
  &=& \min_{y \in \mathbb{R}^d} \VTTNormS{ Xb^{\perp} - XAy}    \\
\label{eqn:ds2}
  &=& \min_{z \in \mathbb{R}^d} \VTTNormS{ Xb^{\perp} - XU_Az }.
\end{eqnarray}
(\ref{eqn:ds1}) follows since $b=Ax_{opt}+b^{\perp}$ and (\ref{eqn:ds2}) follows since the columns of the matrix $A$ span the same subspace as the columns of $U_A$. Now, let $z_{opt} \in \mathbb{R}^d$  be such that $U_A z_{opt} = A(x_{opt}-\tilde{x}_{opt})$, and note that $z_{opt}$ minimizes eqn.~(\ref{eqn:ds2}). The latter fact follows since
$$\VTTNormS{Xb^{\perp} - XA(x_{opt}-\tilde{x}_{opt})} =
\VTTNormS{Xb^{\perp} - X(b-b^{\perp}) +
XA\tilde{x}_{opt}} = \VTTNormS{XA\tilde{x}_{opt}
- Xb}.$$
Thus, by the normal equations~(\ref{eqn:normal_eqn}), we have that
\begin{equation*}
\label{eqn:ds-normal} (XU_A)^TXU_A z_{opt} =
(XU_A)^T Xb^{\perp}.
\end{equation*}
Taking the norm of both sides and observing that under condition~(\ref{eqn:lemma1_ass1}) we have $\sigma_i((XU_A)^TXU_A) = \sigma_i^2(XU_A) \ge 1/\sqrt{2}$, for all $i$, it follows that
\begin{equation}
\label{eqn:z-norm1}
  \VTTNormS{z_{opt}} / 2  \le \VTTNormS{(XU_A)^TXU_Az_{opt}} = \VTTNormS{
(XU_A)^T Xb^{\perp} }.
\end{equation}
Using condition~(\ref{eqn:lemma1_ass2}) we observe that
\begin{equation}
\label{eqn:z-norm2}
   \VTTNormS {z_{opt}} \le \epsilon\mathcal{Z}^2.
\end{equation}

\noindent To establish the first claim of the lemma, let us rewrite
the norm of the residual vector as
\begin{eqnarray}
\VTTNormS{ b - A\tilde{x}_{opt} } \nonumber
   &=& \VTTNormS{ b - Ax_{opt} + Ax_{opt} - A\tilde{x}_{opt} }  \\
\label{eqn:pfCeq1}
   &=& \VTTNormS{ b - Ax_{opt} } + \VTTNormS{ Ax_{opt} - A\tilde{x}_{opt} } \\
\label{eqn:pfCeq2}
   &=& \mathcal{Z}^{2} + \VTTNormS{ U_Az_{opt}} \\
\label{eqn:pfCeq3}
   &\leq& \mathcal{Z}^{2} + \epsilon \mathcal{Z}^{2} ,
\end{eqnarray}
where (\ref{eqn:pfCeq1}) follows by Pythagoras, since $b - Ax_{opt}
= b^\perp$, which is orthogonal to $A$, and consequently to
$A(x_{opt} - \tilde{x}_{opt})$; (\ref{eqn:pfCeq2}) follows
by the definition of $z_{opt}$ and $\mathcal{Z}$; and (\ref{eqn:pfCeq3})
follows by (\ref{eqn:z-norm2}) and the orthogonality of $U_A$. The first claim of the lemma follows since
$\sqrt{1+\epsilon} \le 1+\epsilon$.

To establish the second claim of the lemma, recall that $A(x_{opt}-\tilde{x}_{opt})=U_Az_{opt}$. If we take the norm of both sides of this expression, we have that
\begin{eqnarray}
\VTTNormS{ x_{opt}-\tilde{x}_{opt} } \label{eqn:pfDeq1}
  &\leq& \frac {\VTTNormS{U_Az_{opt}}} {\sigma_{min}^2(A)} \\
\label{eqn:pfDeq2}
  &\leq& \frac {\epsilon\mathcal{Z}^2} {\sigma_{min}^2(A)},
\end{eqnarray}
where (\ref{eqn:pfDeq1}) follows since $\sigma_{min}(A)$ is the
smallest singular value of $A$ and since the rank of $A$ is
$d$; and (\ref{eqn:pfDeq2}) follows by
(\ref{eqn:z-norm2}) and the orthogonality of $U_A$. Taking the square root, the second claim of the
lemma follows.
\end{Proof}

\noindent If we make no assumption on $b$, then~(\ref{eqn:lemma1_eq4}) from
Lemma~\ref{lem:suff_cond} may provide a weak bound in terms of
$\VTTNorm{x_{opt}}$. If, on the other hand, we make the additional
assumption that a constant fraction of the norm of $b$ lies in
the subspace spanned by the columns of $A$,
then~(\ref{eqn:lemma1_eq4}) can be strengthened.
Such an assumption is reasonable, since most least-squares problems are
practically interesting if at least some part of $b$ lies in the subspace
spanned by the columns of $A$.

\begin{lemma}
\label{lem:suff_cond2}
Using the notation of Lemma~\ref{lem:suff_cond} and assuming
that $\VTTNorm{U_AU_A^Tb} \geq \gamma\VTTNorm{b}$, for some fixed
$\gamma \in (0,1]$ it follows that
\begin{equation}
\VTTNorm{x_{opt}-\tilde{x}_{opt}}
  \leq \sqrt{\epsilon}\left(\kappa(A)\sqrt{\gamma^{-2}-1}\right)\VTTNorm{x_{opt}} .
\end{equation}
\end{lemma}
\begin{Proof}
Since $\VTTNorm{U_AU_A^Tb} \geq \gamma\VTTNorm{b}$, it follows that
\begin{eqnarray}
         \mathcal{Z}^2
\nonumber    &=&    \VTTNormS{b} - \VTTNormS{U_A U_A^T b}          \\
\nonumber    &\leq& (\gamma^{-2}-1) \VTTNormS{U_A U_A^T b}         \\
\nonumber    &\leq&
{\sigma_{\max}^{2}(A)}(\gamma^{-2}-1)\VTTNormS{x_{opt}}  .
\end{eqnarray}
This last inequality follows from $U_AU_A^Tb = Ax_{opt}$, which implies $$ \VTTNorm{U_A U_A^T b} = \VTTNorm{Ax_{opt}} \leq \TNorm{A} \VTTNorm{x_{opt}} = \sigma_{\max}\left(A\right)\VTTNorm{x_{opt}}. $$
By combining this with eqn. (\ref{eqn:lemma1_eq4}) of
Lemma~\ref{lem:suff_cond}, the lemma follows.
\end{Proof}

\section{A sampling-based randomized algorithm} \label{sxn:sampling}

In this section, we present our randomized sampling algorithm for the least squares approximation problem of eqn.~(\ref{eqn:orig_ls_prob}). We also state and prove an associated quality-of-approximation theorem.

\subsection{The main algorithm and main theorem} \label{sxn:sampling:result}

Algorithm~\ref{alg:alg_sample_fast} takes as input a matrix $A \in \mathbb{R}^{n \times d}$, a vector $b \in \mathbb{R}^n$, and an error parameter $\epsilon \in (0,1)$. This algorithm starts by preprocessing the matrix $A$ and the vector $b$ with the Randomized Hadamard Transform. It then constructs a smaller problem by sampling uniformly at random a small number of constraints from the preprocessed problem. Our main quality-of-approximation theorem (Theorem~\ref{thm:alg_sample_fast} below) states that with constant probability over the random choices made by the algorithm, the vector
$\tilde{x}_{opt}$ returned by this algorithm will satisfy the relative-error bounds of eqns.~(\ref{eqn:result1_intro}) and~(\ref{eqn:result2_intro}) and will be computed quickly.

\begin{algorithm}[h]
\begin{framed}

\textbf{Input:} $A \in \mathbb{R}^{n \times d}$, $b \in
\mathbb{R}^n$, and an error parameter $\epsilon \in (0,1)$.

\vspace{0.1in}

\textbf{Output:} $\tilde{x}_{opt} \in \mathbb{R}^d$.

\begin{enumerate}

\item Let $r$ assume the value of eqn.~(\ref{eqn:rvaluefinal}).

\item Let $S$ be an empty matrix.

\item \textbf{For} $t=1,\ldots,r$ (i.i.d. trials with replacement) \textbf{select uniformly at random} an integer from $\left\{1,2,\ldots,n\right\}$.

\begin{itemize}

\item \textbf{If} $i$ is selected, \textbf{then} append the column vector $\left(\sqrt{n/r}\right) e_i$ to $S$, where $e_i \in \mathbb{R}^n$ is an all-zeros vector except for its $i$-th entry which is set to one.

\end{itemize}

\item Let $H \in \mathbb{R}^{n\times n}$ be the normalized Hadamard transform
matrix.

\item Let $D \in \mathbb{R}^{n \times n}$ be a diagonal matrix with
$$
D_{ii}
            = \left\{ \begin{array}{ll}
                         +1 & \mbox{, with probability $1/2$} \\
                         -1 & \mbox{, with probability $1/2$} \\
                      \end{array}
              \right.
$$
\item
Compute and return $\tilde{x}_{opt} = \left(S^THDA\right)^\dagger S^THDb $.
\end{enumerate}

\end{framed}
\caption{A fast random sampling algorithm for least squares
approximation} \label{alg:alg_sample_fast}
\end{algorithm}

In more detail, after preprocessing with the Randomized Hadamard Transform of Section~\ref{sxn:RHT}, Algorithm~\ref{alg:alg_sample_fast} samples exactly $r$ constraints from the preprocessed least squares problem, rescales each sampled constraint by $\sqrt{n/r}$, and solves the least squares problem induced on just those sampled and rescaled constraints. (Note that the algorithm explicitly computes only those rows of $HDA$ and only those elements of $HDb$ that need to be accessed.) More formally, we will let $S \in \mathbb{R}^{n \times r}$ denote a sampling matrix specifying which of the $n$ constraints are to be sampled and how they are to be rescaled. This matrix is initially empty and is constructed as described in Algorithm~\ref{alg:alg_sample_fast}. Then, we can consider the problem
\begin{equation*}
\tilde{\mathcal{Z}}
   = \min_{x \in \mathbb{R}^d} \VTTNorm{S^THDAx- S^THDb } ,
\end{equation*}
which is just a least squares approximation problem involving the $r$ constraints sampled from the matrix $A$ after the preprocessing with the Randomized Hadamard Transform. The minimum $\ell_2$-norm vector $\tilde{x}_{opt} \in \mathbb{R}^d$ among those that achieve the minimum value $\tilde{\mathcal{Z}}$ in this problem is
\begin{equation*}
\tilde{x}_{opt}
   = \left(S^THDA\right)^{\dagger}S^THDb     ,
\end{equation*}
which is the output of Algorithm~\ref{alg:alg_sample_fast}.
\begin{theorem}
\label{thm:alg_sample_fast}
Suppose $A \in \mathbb{R}^{n \times d}$, $b \in \mathbb{R}^{n}$, and let $\epsilon \in (0,1)$. Run Algorithm~\ref{alg:alg_sample_fast} with
\begin{equation}\label{eqn:rvaluefinal}
r = \max\left\{48^2 d \ln\left(40nd\right)\ln\left(100^2d \ln \left(40nd\right)\right),
40d\ln(40nd)/\epsilon\right\}
\end{equation}
and return $\tilde{x}_{opt}$. Then, with probability at least .8, the following two claims hold: first, $\tilde{x}_{opt}$ satisfies
$$
\VTTNorm{A\tilde{x}_{opt}-b} \le (1+\epsilon) \mathcal{Z};
$$
and, second, if we assume that $\VTTNorm{U_A U_A^T b} \ge \gamma \VTTNorm{b}$ for some $\gamma \in (0,1]$, then $\tilde{x}_{opt}$ satisfies
$$
\VTTNorm{x_{opt}-\tilde{x}_{opt}}
  \leq \sqrt{\epsilon}\left(\kappa(A)\sqrt{\gamma^{-2}-1}\right)\VTTNorm{x_{opt}}.
$$
Finally,
$$ n(d+1) + 2n(d+1) \log_2 \left(r + 1\right) + O\left(rd^2 \right)$$
time suffices to compute the solution $\tilde{x}_{opt}$.
\end{theorem}
\noindent \textbf{Remark:} Assuming that $d \leq n \leq e^d$, and using $\max\{a_1,a_2\} \leq a_1 + a_2$, we get that $$r = O\left(d(\ln d)(\ln n) + \frac{d \ln n}{\epsilon}\right).$$ Thus, the running time of Algorithm~\ref{alg:alg_sample_fast} becomes
$$O\left(nd\ln \frac{d}{\epsilon} + d^3 (\ln d)(\ln n) + \frac{d^3 \ln n}{\epsilon}\right).$$
Assuming that $\frac{n}{\ln n} = \Omega(d^2)$, the above running time reduces to $$O\left(nd \ln \frac{d}{\epsilon} + \frac{nd \ln d}{\epsilon}\right).$$ It is worth noting that improvements over the standard $O(nd^2)$ time could be derived with weaker assumptions on $n$ and $d$. However, for the sake of clarity of presentation, we only focus on the above setting.

\noindent \textbf{Remark:}
The assumptions in our theorem have a natural geometric interpretation.%
\footnote{We would like to thank Ilse Ipsen for pointing out to us this
geometric interpretation.}
In particular, they imply that our approximation becomes worse as the angle
between the vector $b$ and the column space of $A$ increases.
To see this, let $\mathcal{Z} = ||Ax_{opt}-b||_2$, and note that
$||b||_2^2 = ||U_AU^T_A b||_2^2 + \mathcal{Z}^2$.
Hence the assumption
$||U_AU^T_kb||_2 \ge \gamma ||b||_2$ can be simply stated as
$$\mathcal{Z} \le \sqrt{1-\gamma^2}||b||_2.$$
The fraction $\mathcal{Z}/||b||_2$ is the sine of the angle between $b$ and
the column space of $A$; see page 242 of~\cite{GVL96}.
Thus, $\sqrt{\gamma^{-2}-1}$ is a bound on the tangent between $b$ and the
column space of $A$; see page 244 of~\cite{GVL96}.
This means that the bound for $||x_{opt}-\tilde{x}_{opt}||_2$ is
proportional to this tangent.

\subsection{The effect of the Randomized Hadamard Transform}
\label{sxn:review_previous:hadamard}

In this subsection, we state a lemma that quantifies the manner in which $HD$ approximately ``uniformizes'' information in the left singular subspace of the matrix $A$. We state the lemma for a general $n \times d$ orthogonal matrix $U$ such that $U^TU = I_d$, although we will be interested in the case when $n \gg d$ and $U$ consists of the top $d$ left singular vectors of the matrix $A$.

\begin{lemma} \label{lem:HU}
Let $U$ be an $n \times d$ orthogonal matrix and let the product $HD$ be the $n \times n$ Randomized Hadamard Transform of Section~\ref{sxn:RHT}. Then, with probability at least $.95$,
\begin{eqnarray}
\label{eqn:lem:HU_eqn2} \VTTNormS{\left(HDU\right)_{(i)}}
   &\leq& \frac{2d\ln(40nd)}{n},\qquad
       \text{ for all } i \in [n]   .
\end{eqnarray}
\end{lemma}
\begin{Proof}
We follow the proof of Lemma 2.1 in~\cite{AC06}. In that lemma, the authors essentially prove that the Randomized Hadamard Transform $HD$ ``spreads out'' input vectors. More specifically, since the columns of the matrix $U$ (denoted by $U^{(j)}$ for all $j \in [d]$) are unit vectors, they prove that for fixed $j \in [d]$ and fixed $i \in [n]$,
$$\Probab{\abs{\left(HDU^{(j)}\right)_i}\geq s}\leq 2 e^{-s^2 n/2}.$$
(Note that we consider $d$ vectors in $\mathbb{R}^n$ whereas~\cite{AC06} considered $n$ vectors in $\mathbb{R}^d$ and thus the roles of $n$ and $d$ are inverted in our proof.) Let $s = \sqrt{2n^{-1}\ln(40nd)}$ to get
$$\Probab{\abs{\left(HDU^{(j)}\right)_i}\geq \sqrt{2n^{-1}\ln(40nd)}}\leq \frac{1}{20nd}.$$
From a standard union bound, this immediately implies that with probability at least $1-1/20$,
\begin{equation}\label{eqn:eqpd12}
\abs{\left(HDU^{(j)}\right)_i}\leq \sqrt{2n^{-1}\ln(40nd)}
\end{equation}
holds for all $i \in [n]$ and $j \in [d]$ . Using
\begin{equation}\label{eqn:eqP1}
\VTTNormS{\left(HDU\right)_{(i)}} = \sum_{j=1}^d \left(HDU^{(j)}\right)_i^2 \leq \frac{2d\ln(40nd)}{n}
\end{equation}
for all $i \in [n]$, we conclude the proof of the lemma.
\end{Proof}

\subsection{Satisfying condition~(\ref{eqn:lemma1_ass1})}

We now establish the following lemma which states that all the singular values of $S^THDU_A$ are close to one. The proof of Lemma~\ref{lem:sample_lem20pf} depends on a bound for approximating the product of a matrix times its transpose by sampling (and rescaling) a small number of columns of the matrix. This bound appears as Theorem~\ref{thm:theorem7correct} in the Appendix and is an improvement over prior work of ours in~\cite{DMM08_CURtheory_JRNL}.

\begin{lemma}
\label{lem:sample_lem20pf}
Assume that eqn.~(\ref{eqn:lem:HU_eqn2}) holds. If
\begin{equation}\label{eqn:rvalue}
r \geq 48^2 d \ln\left(40nd\right)\ln\left(100^2d \ln \left(40nd\right)\right)
\end{equation}
then, with probability at least .95,
$$\abs{1 - \sigma_i^2\left(S^THDU_A\right)} \leq 1-\frac{1}{\sqrt{2}}, $$
holds for all $i \in [d]$.
\end{lemma}
\begin{Proof}
Note that for all $i \in [d]$
\begin{eqnarray}
\abs{1 - \sigma_i^2\left(S^THDU_A\right)} \nonumber
   &=&    \abs{\sigma_i\left(U_A^TDH^THDU_A\right)
            - \sigma_i\left(U_A^TDH^TSS^THDU_A\right)}    \\
\label{eqn:eqX31}
   &\leq& \TNorm{U_A^TDH^THDU_A - U_A^TDH^TSS^THDU_A}.
\end{eqnarray}
In the above, we used the fact that $U_A^TDH^THD U_A = I_d$. We now can view $U_A^TDSS^TH^THDU_A$ as an approximation to the product of two matrices $U_A^TDH^T=\left(HDU_A\right)^T$ and $HDU_A$ by randomly sampling and rescaling columns of $\left(HDU_A\right)^T$. Thus, we can leverage Theorem~\ref{thm:theorem7correct} from the Appendix. More specifically, consider the matrix $\left(HDU_A\right)^T$. Obviously, since $H$, $D$, and $U_A$ are orthogonal matrices, $\TNorm{HDU_A}=1$ and $\FNorm{HDU_A}=\FNorm{U_A}=\sqrt{d}$. Let $\beta = \left(2\ln(40nd)\right)^{-1}$; since we assumed that eqn.~(\ref{eqn:lem:HU_eqn2}) holds, we note that the columns of $\left(HDU_A\right)^T$, which correspond to the rows of $HDU_A$, satisfy
\begin{equation}
\label{eqn:unif_prob_OK} \frac{1}{n}
   \ge  \beta \frac{\VTTNormS{\left(HDU_A\right)_{(i)}}}{\FNormS{HDU_A}}, \qquad
       \text{ for all } i \in [n]   .
\end{equation}
Thus, applying Theorem~\ref{thm:theorem7correct} with $\beta$ as above, $\epsilon = 1 - \left(1/\sqrt{2}\right)$, and $\delta = 1/20$ implies that
$$\TNorm{U_A^TDH^THU_A - U_A^TDH^TSS^THDU_A} \leq 1-\frac{1}{\sqrt{2}} $$
holds with probability at least $1-1/20=.95$. For the above bound to hold, we need $r$ to assume the value of eqn.~(\ref{eqn:rvalue}). Finally, we note that since $\FNormS{HDU_A}=d \geq 1$, the assumption of Theorem~\ref{thm:theorem7correct} on the Frobenius norm of the input matrix is always satisfied. Combining the above with inequality~(\ref{eqn:eqX31}) concludes the proof of the lemma.
\end{Proof}

\subsection{Satisfying condition~(\ref{eqn:lemma1_ass2})}

We next prove the following lemma, from which it will follow that condition~(\ref{eqn:lemma1_ass2}) is satisfied by Algorithm~\ref{alg:alg_sample_fast}. The proof of this lemma depends on bounds for randomized matrix multiplication algorithms that appeared in~\cite{dkm_matrix1}.

\begin{lemma}
\label{lem:sample_lem40pf}
If eqn.~(\ref{eqn:lem:HU_eqn2}) holds and $r \geq 40d\ln(40nd)/\epsilon$, then with probability at least .9,
$$\VTTNormS{
\left(S^THDU_A\right)^{T}S^THDb^{\perp}
} \leq \epsilon \mathcal{Z}^{2}/2.$$
\end{lemma}
\begin{Proof}
Recall that $b^{\perp} = U_A^{\perp}{U_A^{\perp}}^{T}b$ and that ${\cal Z} = \VTTNorm{b^{\perp}}$. We start by noting that since
$\VTTNormS{U_A^TDH^THDb^{\perp}}=\VTTNormS{U_A^Tb^{\perp}}=0$ it follows that
$$
 \VTTNormS{ \left(S^THDU_A\right)^{T}S^THDb^{\perp} }
   = \VTTNormS{U_A^TDH^TSS^THDb^{\perp}
             - U_A^TDH^THDb^{\perp}}    .
$$
Thus, we can view $\left(S^THDU_A\right)^{T}S^THDb^{\perp}$ as approximating the product of two matrices $\left(HDU_A\right)^{T}$ and $HDb^{\perp}$ by randomly sampling columns from $\left(HDU_A\right)^T$ and rows/elements from $HDb^{\perp}$. Note that the sampling probabilities are uniform and do not depend on the norms of the columns of $\left(HDU_A\right)^{T}$ or the rows of $\mathcal{H}b^{\perp}$. However, we can still apply the results of Table 1 (second row) in page 150 of~\cite{dkm_matrix1}. More specifically, since we condition on eqn.~(\ref{eqn:lem:HU_eqn2}) holding, the rows of $HDU_A$ (which of course correspond to columns of $\left(HDU_A\right)^T$) satisfy
\begin{equation}
\frac{1}{n}   \ge  \beta \frac{\VTTNormS{\left(HDU_A\right)_{(i)}}}{\FNormS{HDU_A}}, \qquad       \text{ for all } i \in [n],
\end{equation}
for $\beta = \left(2\ln(40nd)\right)^{-1}$. Applying the result of Table 1 (second row) of~\cite{dkm_matrix1} we get
\begin{equation*}
\Expect{ \VTTNormS{\left(S^THDU_A\right)^{T}S^THDb^{\perp}
} }
   \leq \frac{1}{\beta r}\FNormS{HDU_A}\VTTNormS{HDb^{\perp}}
   =     \frac{d{\cal Z}^2}{\beta r}.
\end{equation*}
In the above we used $\FNormS{HDU_A} = d$. Markov's inequality now implies that with probability at least .9,
\begin{equation*}
\VTTNormS{\left(S^THDU_A\right)^{T}S^THDb^{\perp}
} \leq \frac{10d{\cal Z}^2}{\beta r}.
\end{equation*}
Setting $r \geq 20\beta^{-1}d/\epsilon$ and using the value of $\beta$ specified above concludes the proof of the lemma.
\end{Proof}

\subsection{Completing the proof of Theorem~\ref{thm:alg_sample_fast}}
\label{sxn:sampling:proofs:complete}

We now complete the proof of Theorem~\ref{thm:alg_sample_fast}. First, let ${\cal E}_{(\ref{eqn:lem:HU_eqn2})}$ denote the event that eqn.~(\ref{eqn:lem:HU_eqn2}) holds; clearly, $\Probab{{\cal E}_{(\ref{eqn:lem:HU_eqn2})}} \geq .95$. Second, let ${\cal E}_{\ref{lem:sample_lem20pf},\ref{lem:sample_lem40pf}|(\ref{eqn:lem:HU_eqn2})}$ denote the event that both Lemmas~\ref{lem:sample_lem20pf} and~\ref{lem:sample_lem40pf} hold conditioned on ${\cal E}_{(\ref{eqn:lem:HU_eqn2})}$ holding. Then,
\begin{eqnarray*}
{\cal E}_{\ref{lem:sample_lem20pf},\ref{lem:sample_lem40pf}|(\ref{eqn:lem:HU_eqn2})} &=& 1 - \overline{{\cal E}_{\ref{lem:sample_lem20pf},\ref{lem:sample_lem40pf}|(\ref{eqn:lem:HU_eqn2})}}\\
&=& 1 - \Probab{\left(\mbox{Lemma \ref{lem:sample_lem20pf} does not hold $|$ \cal E}_{(\ref{eqn:lem:HU_eqn2})}\right) \textbf{OR} \left(\mbox{Lemma \ref{lem:sample_lem40pf} does not hold $|$ \cal E}_{(\ref{eqn:lem:HU_eqn2})}\right)}\\
&\geq& 1 - \Probab{\mbox{Lemma \ref{lem:sample_lem20pf} does not hold $|$ \cal E}_{(\ref{eqn:lem:HU_eqn2})}} -\Probab{\mbox{Lemma \ref{lem:sample_lem40pf} does not hold $|$ \cal E}_{(\ref{eqn:lem:HU_eqn2})}}\\
&\geq& 1 - .05 - .1 = .85.
\end{eqnarray*}
In the above, $\overline{\cal E}$ denotes the complement of event ${\cal E}$. In the first inequality we used the union bound and in the second inequality we leveraged the bounds for the failure probabilities of Lemmas~\ref{lem:sample_lem20pf} and~\ref{lem:sample_lem40pf} given that eqn.~(\ref{eqn:lem:HU_eqn2}) holds. We now let ${\cal E}$ denote the event that both Lemmas~\ref{lem:sample_lem20pf} and~\ref{lem:sample_lem40pf} hold, without any a priori conditioning on event ${\cal E}_{(\ref{eqn:lem:HU_eqn2})}$; we will bound $\Probab{\cal E}$ as follows:
\begin{eqnarray*}
\Probab{\cal E} &=& \Probab{{\cal E} | {\cal E}_{(\ref{eqn:lem:HU_eqn2})}}\cdot \Probab{{\cal E}_{(\ref{eqn:lem:HU_eqn2})}}
+\Probab{{\cal E} | \overline{{\cal E}_{(\ref{eqn:lem:HU_eqn2})}}}\cdot \Probab{\overline{{\cal E}_{(\ref{eqn:lem:HU_eqn2})}}}\\
&\geq& \Probab{{\cal E} | {\cal E}_{(\ref{eqn:lem:HU_eqn2})}}\cdot \Probab{{\cal E}_{(\ref{eqn:lem:HU_eqn2})}}\\
&=& \Probab{{\cal E}_{\ref{lem:sample_lem20pf},\ref{lem:sample_lem40pf}|(\ref{eqn:lem:HU_eqn2})} | {\cal E}_{(\ref{eqn:lem:HU_eqn2})}}\cdot \Probab{{\cal E}_{(\ref{eqn:lem:HU_eqn2})}}\\
&\geq& .85\cdot .95 \geq .8.
\end{eqnarray*}
In the first inequality we used the fact that all probabilities are positive. The above derivation immediately bounds the success probability of Theorem~\ref{thm:alg_sample_fast}. Combining Lemmas~\ref{lem:sample_lem20pf} and~\ref{lem:sample_lem40pf} with the structural results of Lemma~\ref{lem:suff_cond} and setting $r$ as in eqn.~(\ref{eqn:rvaluefinal}) concludes the proof of the accuracy guarantees of Theorem~\ref{thm:alg_sample_fast}.

We now discuss the running time of Algorithm~\ref{alg:alg_sample_fast}. First of all, by the construction of $S$, the number of non-zero entries in $S$ is $r$. In Step $6$ we need to compute the products $S^THDA$ and $S^THDb$. Recall that $A$ has $d$ columns and thus the running time of computing both products is equal to the time needed to apply $S^THD$ on $(d+1)$ vectors. First, note that in order to apply $D$ on $(d+1)$ vectors in $\mathbb{R}^n$, $n(d+1)$ operations suffice. In order to estimate how many operations are needed to apply $S^TH$ on $(d+1)$ vectors, we use the results of Theorem~$2.1$ (see also Section 7) of Ailon and Liberty~\cite{AL08}, which state that at most $2n(d+1)\log_2 \left(\abs{S}+1\right)$ operations are needed for this operation. Here $\abs{S}$ denotes the number of non-zero elements in the matrix $S$, which is at most $r$. After this preprocessing, Algorithm~\ref{alg:alg_sample_fast} must compute the pseudoinverse of an $r \times d$ matrix, or, equivalently, solve a least-squares problem on $r$ constraints and $d$ variables. This operation can be performed in $O(rd^2)$ time since $r \geq d$. Thus, the entire algorithm runs in time
$$n(d+1) + 2n(d+1) \log_2 \left(r + 1\right) + O\left(rd^2 \right).$$
%

\section{A projection-based randomized algorithm}
\label{sxn:projection}

In this section, we present a projection-based randomized algorithm for the least squares approximation problem of eqn. (\ref{eqn:orig_ls_prob}). We also state and prove an associated quality-of-approximation theorem.

\subsection{The main algorithm and main theorem}
\label{sxn:projection:result}

Algorithm~\ref{alg:alg_projection_fast} takes as input a matrix $A \in \mathbb{R}^{n \times d}$, a vector $b \in \mathbb{R}^n$, and an error parameter $\epsilon \in (0,1/2)$. This algorithm also starts by preprocessing the matrix $A$ and right hand side vector $b$ with the Randomized Hadamard Transform. It then constructs a smaller problem by performing a ``sparse projection'' on the preprocessed problem. Our main quality-of-approximation theorem (Theorem~\ref{thm:alg_projection_fast} below) will state that with constant probability (over the random choices made by the algorithm) the vector
$\tilde{x}_{opt}$ returned by this algorithm will satisfy the relative-error bounds of eqns.~(\ref{eqn:result1_intro}) and~(\ref{eqn:result2_intro}) and will be computed quickly.

\begin{algorithm}[h]
\begin{framed}

\textbf{Input:} $A \in \mathbb{R}^{n \times d}$, $b \in \mathbb{R}^n$, and an error parameter $\epsilon \in (0,1/2)$.

\vspace{0.1in}

\textbf{Output:} $\tilde{x}_{opt} \in \mathbb{R}^d$.

\begin{enumerate}

\item Let $q$ and $k$ assume the values of eqns.~(\ref{eqn:valueqfinal}) and~(\ref{eqn:valuekfinal}).

\item Let $T \in \mathbb{R}^{k \times n}$ be a random matrix with
$$
T_{ij}
            = \left\{\begin{array}{ll}
                        +\sqrt{\frac{1}{kq}}& \mbox{, with probability $q/2$} \\
                        -\sqrt{\frac{1}{kq}}& \mbox{, with probability $q/2$} \\
                        0                   & \mbox{, with probability $1-q$,}
                     \end{array}
              \right.
$$
for all $i,j$ independently.
\item Let $H \in \mathbb{R}^{n\times n}$ be the normalized Hadamard transform
matrix.

\item Let $D \in \mathbb{R}^{n \times n}$ be a diagonal matrix with
$$
D_{ii}
            = \left\{ \begin{array}{ll}
                         +1 & \mbox{, with probability $1/2$} \\
                         -1 & \mbox{, with probability $1/2$} \\
                      \end{array}
              \right.
$$
\item
Compute and return
$\tilde{x}_{opt}
   = \left(THDA\right)^{\dagger}THDb$.
\end{enumerate}

\end{framed}
\caption{A fast random projection algorithm for least squares
approximation} \label{alg:alg_projection_fast}
\end{algorithm}

In more detail, Algorithm~\ref{alg:alg_projection_fast} begins by preprocessing the matrix $A$ and right hand side vector $b$ with the Randomized Hadamard Transform $HD$ of Section~\ref{sxn:RHT}. This algorithm explicitly computes only those rows of $HDA$ and those elements of $HDb$ that need to be accessed to perform the sparse projection. After this initial preprocessing, Algorithm~\ref{alg:alg_projection_fast} will perform a ``sparse projection'' by multiplying $HDA$ and $HDb$ by the sparse matrix $T$ (described in more detail in
Section~\ref{sxn:review_previous:projecting}). Then, we can consider the problem
\begin{equation*}
\tilde{\mathcal{Z}}
   = \min_{x \in \mathbb{R}^d} \VTTNorm{THDAx- THDb } ,
\end{equation*}
which is just a least squares approximation problem involving the matrix $THDA \in \mathbb{R}^{k \times d}$ and the vector $THDb \in \mathbb{R}^k$. The minimum $\ell_2$-norm vector $\tilde{x}_{opt} \in \mathbb{R}^d$ among those that achieve the minimum value $\tilde{\mathcal{Z}}$ in this problem is
\begin{equation*}
\tilde{x}_{opt} = \left(THDA\right)^{\dagger}THDb,
\end{equation*}
which is the output of Algorithm~\ref{alg:alg_projection_fast}.
\begin{theorem} \label{thm:alg_projection_fast}
Suppose $A \in \mathbb{R}^{n \times d}$, $b \in \mathbb{R}^{n}$, and let $\epsilon \in (0,1/2)$. Run Algorithm~\ref{alg:alg_projection_fast} with\footnote{$C_q$ and $C_k$ are the unspecified constants of Lemma~\ref{lem:matousek}.}
\begin{eqnarray}
\label{eqn:valueqfinal} q &\geq& \frac{C_q d \ln(40nd)}{n}\left(2\ln n + 16d + 16\right)\\
\label{eqn:valuekfinal} k &\geq& \max\left\{C_k \left(118^2 d + 98^2\right),\frac{60d}{\epsilon}\right\}
\end{eqnarray}
and return $\tilde{x}_{opt}$. Then, with probability at least $.8$, the following two claims hold:
first, $\tilde{x}_{opt}$ satisfies
$$
\VTTNorm{A\tilde{x}_{opt}-b} \le (1+\epsilon) \mathcal{Z}  ;
$$
and, second, if we assume that
$\VTTNorm{U_A U_A^T b} \ge \gamma \VTTNorm{b}$ for some $\gamma \in (0,1]$
then $\tilde{x}_{opt}$ satisfies
$$
\VTTNorm{x_{opt}-\tilde{x}_{opt}}
  \leq \sqrt{\epsilon}\left(\kappa(A)\sqrt{\gamma^{-2}-1}\right)\VTTNorm{x_{opt}}.
$$
Finally, the expected running time of the algorithm is (at most)
$$
n(d+1) + 2n(d+1) \log_2 \left(nkq + 1\right) + O\left(kd^2 \right).
$$
\end{theorem}
\noindent \textbf{Remark:} Assuming that $d \leq n \leq e^d$ we get that
$$
q = O\left(\frac{d^2 \ln n}{n}\right)\qquad \text{and}\qquad
k = O\left(\frac{d}{\epsilon}\right).
$$
Thus, the expected running time of Algorithm~\ref{alg:alg_projection_fast} becomes
$$O\left(nd\ln \frac{d}{\epsilon} + \frac{d^3}{\epsilon}\right).$$
Finally, assuming $n = \Omega(d^2)$, the above running time reduces to $$O\left(nd\ln \frac{d}{\epsilon}+\frac{nd}{\epsilon}\right).$$ It is worth noting that improvements over the standard $O(nd^2)$ time could be derived with weaker assumptions on $n$ and $d$.

\subsection{Sparse projection matrices}
\label{sxn:review_previous:projecting}

In this subsection, we state a lemma about the action of a sparse random matrix operating on a vector. Recall that given any set of $n$ points in Euclidean space, the Johnson-Lindenstrauss lemma states that those points can be mapped via a linear function to $k=O(\epsilon^{-2}\ln n)$ dimensions such that the distances between all pairs of points are preserved to within a multiplicative factor of $1 \pm \epsilon$; see~\cite{Matousek08_RSA} and references therein for details.

Formally, let $\epsilon \in (0,1/2)$ be an error parameter, $\delta \in (0,1)$ be a failure probability, and $\alpha \in [1/\sqrt{n},1]$ be a ``uniformity'' parameter. In addition, let $q$ be a ``sparsity'' parameter defining the expected number of nonzero elements per row, and let $k$ be the number of rows in our matrix. Then, define the $k \times n$ random matrix $T$ as in Algorithm
\ref{alg:alg_projection_fast}. Matou\v{s}ek proved the following lemma, as the key step in his version of the Ailon-Chazelle result~\cite{AC06,Matousek08_RSA}.

\begin{lemma}\label{lem:matousek}
Let $T$ be the sparse random matrix of Algorithm~\ref{alg:alg_projection_fast}, where $q = C_q \alpha^2\ln(\frac{n}{\epsilon\delta})$ for some sufficiently large constant $C_q$ (but still such that $q \le 1$), and $k = C_k \epsilon^{-2} \ln(\frac{4}{\delta}$) for some sufficiently large constant $C_k$ (but such that $k$ is integral). Then for every vector $x \in \mathbb{R}^{n}$ such that $\VINorm{x}/\VTTNorm{x} \le \alpha$, we have that with probability at least $1-\delta$
$$
\abs{\VTTNorm{Tx}-\VTTNorm{x}} \le \epsilon \VTTNorm{x}.
$$
\end{lemma}

\noindent \textbf{Remark:} In order to achieve sufficient concentration for all vectors $x \in \mathbb{R}^{n}$, the linear mapping defining the Johnson-Lindenstrauss transform is typically
``dense,'' in the sense that almost all the elements in each of the $k$ rows of the matrix defining the mapping are nonzero. In this case, implementing the mapping on $d$ vectors (in, e.g., a matrix $A$) via a matrix multiplication requires $O(ndk)$ time. This is not faster than the $O(nd^2)$ time required to compute an exact solution to the problem of eqn.~(\ref{eqn:orig_ls_prob}) if $k$ is at least $d$. The Ailon-Chazelle result~\cite{AC06,Matousek08_RSA} states that the mapping can be ``sparse,'' in the sense that only a few of the elements in each of the $k$ rows need to be nonzero, provided that the vector $x$ is ``well-spread,'' in the sense that $\VINorm{x}/\VTTNorm{x}$ is close to $1/\sqrt{n}$. This is exactly what the preprocessing with the Randomized Hadamard Transform guarantees.

\subsection{Proof of Theorem~\ref{thm:alg_projection_fast}}
\label{sxn:projection:proofs}

In this subsection, we provide a proof of Theorem~\ref{thm:alg_projection_fast}. Recall that by the results of Section~\ref{sxn:sampling:proofs:structural}, in order to prove Theorem~\ref{thm:alg_projection_fast}, we must show that the matrix $THD$ constructed by Algorithm~\ref{alg:alg_projection_fast} satisfies conditions~(\ref{eqn:lemma1_ass1}) and~(\ref{eqn:lemma1_ass2}) with probability at least $.5$. The next two subsections focus on proving that these conditions hold; the last subsection discusses the running time of Algorithm~\ref{alg:alg_projection_fast}.

\subsubsection{Satisfying condition~(\ref{eqn:lemma1_ass1})}
In order to prove that all the singular values of $THDU_A$ are close to one, we start with the following lemma which provides a means to bound the spectral norm of a matrix. This lemma is an instantiation of lemmas that appeared in~\cite{AHK06,FO05}.
\begin{lemma}\label{lem:AroraHazenKale}
Let $M$ be a $d \times d$ symmetric matrix and define the grid
\begin{equation} \label{eqn:grid}
\Omega=\left\{x: x\in\frac{1}{2\sqrt{d}}\mathbb{Z}^{d},\VTTNorm{x}\le1\right\}  .
\end{equation}
In words, $\Omega$ includes all $d$-dimensional vectors $x$ whose coordinates are integer multiples of $\left(2\sqrt{d}\right)^{-1}$ and satisfy $\VTTNorm{x}\leq 1$. Then, the cardinality of $\Omega$ is at most $e^{4d}$. In addition, if for every $x,y\in\Omega$ we have that $\abs{x^TMy}\le \epsilon'$, then for every unit vector $x$ we have that $\abs{x^TMx}\le 4\epsilon'$.
\end{lemma}
We next establish Lemma~\ref{lem:project_lem20pf}, which states that all the singular values of $THDU_A$ are close to one with constant probability. The proof of this lemma depends on the bound provided by Lemma~\ref{lem:AroraHazenKale} and it immediately shows that condition~(\ref{eqn:lemma1_ass1}) is satisfied by Algorithm~\ref{alg:alg_projection_fast}.
\begin{lemma}\label{lem:project_lem20pf}
Assume that Lemma~\ref{lem:HU} holds. If $q$ and $k$ satisfy:
\begin{eqnarray}
\label{eqn:chooseq} q &\geq& \frac{C_q d \ln(40nd)}{n}\left(2\ln n + 16d + 16\right)\\
\label{eqn:choosek} k &\geq& C_k \left(118^2 d + 98^2\right),
\end{eqnarray}
then, with probability at least .95,
$$\abs{1 - \sigma_i^2\left(THDU_A\right)} \leq 1-(1/\sqrt{2})$$
holds for all $i \in [d]$. Here $C_q$ and $C_k$ are the unspecified constants of Lemma~\ref{lem:matousek}.
\end{lemma}
\begin{Proof}
Define the symmetric matrix $M=U_A^TDH^TT^TTHDU_A-I_d \in \mathbb{R}^{d\times d}$, recall that $I_d=U_A^TDH^THDU_A$, and note that
\begin{eqnarray} \label{eqn:eqX32}
\abs{1-\sigma_i^2\left(THDU_A\right)} \leq \TNorm{M}
\end{eqnarray}
holds for all $i \in [d]$. Consider the grid $\Omega$ of eqn.~(\ref{eqn:grid}) and note that there are no more than $e^{8d}$ pairs $(x,y)\in\Omega\times\Omega$, since $\abs{\Omega} \le
e^{4d}$ by Lemma~\ref{lem:AroraHazenKale}. Since $\TNorm{M} = \sup_{\VTTNorm{x}=1}\abs{x^TMx}$, in order to show that $\TNorm{M}\le 1-2^{-1/2}$, it suffices by Lemma~\ref{lem:AroraHazenKale} to show that $\abs{x^TMy}\le \left(1-2^{-1/2}\right)/4$, for all $x,y\in\Omega$. To do so, first, consider a single $x,y$ pair. Let
\begin{eqnarray*}
\Delta_1 &=& \VTTNormS{THDU_A(x+y)}-\VTTNormS{HDU_A(x+y)}  \\
\Delta_2 &=& \VTTNormS{THDU_Ax}-\VTTNormS{HDU_Ax} \\
\Delta_3 &=&\VTTNormS{THDU_Ay}-\VTTNormS{HDU_Ay}  ,
\end{eqnarray*}
and note that
\begin{eqnarray*}
\Delta_1 &=& (x+y)^TU_A^TDH^TT^T THDU_A(x+y)-(x+y)^T(x+y).
\end{eqnarray*}
By multiplying out the right hand side of the above equation and rearranging terms, it follows that
\begin{equation}\label{eqn:pd1}
x^TMy = x^TU_A^TDH^TT^TT HD U_Ay-x^Ty
   = \frac{1}{2}\left( \Delta_1 + \Delta_2 + \Delta_3 \right)   .
\end{equation}
In order to use Lemma~\ref{lem:matousek} to bound the quantities $\Delta_1, \Delta_2$, and $\Delta_3$, we need a bound on the uniformity ratio $\VINorm{HDU_Ax}/\VTTNorm{HDU_Ax}$. To do so, note that
$$
\frac{ \VINorm{HDU_Ax} }{ \VTTNorm{HDU_Ax}}= \frac{ \max_{i \in [n]} \abs{\left(HDU_A\right)_{(i)}x} }{\VTTNorm{HDU_Ax} }
      \le \frac{\max_{i \in [n]} \VTTNorm{\left(HDU_A\right)_{(i)}}\VTTNorm{x}}{\VTTNorm{x}}
      \le \sqrt{\frac{2d \ln(40nd)}{n}}.
$$
The above inequalities follow by $\VTTNorm{HDU_Ax}=\VTTNorm{x}$ and Lemma~\ref{lem:HU}. This holds for both our chosen points $x$ and $y$ and in fact for all $x\in\Omega$. Let $\epsilon_1=3/125$ and let $\delta=1/(60e^{8d})$ (these choices will be explained shortly). Then, it follows from Lemma~\ref{lem:matousek} that by setting $\alpha = \sqrt{2d \ln(40nd)/n}$ and our choices for $k$ and $q$, each of the following three statements holds with probability at least $1-\delta$:
\begin{eqnarray*}
\abs{\Delta_1} &\le& \epsilon_1 \VTTNormS{HDU_A(x+y)} = \epsilon_1 \VTTNormS{x+y}\leq 4\epsilon_1  \\
\abs{\Delta_2} &\le& \epsilon_1 \VTTNormS{HDU_Ax} = \epsilon_1\VTTNorm{x}    \le  \epsilon_1  \\
\abs{\Delta_3} &\le& \epsilon_1 \VTTNormS{HDU_Ay}  = \epsilon_1\VTTNorm{y}   \le  \epsilon_1  .
\end{eqnarray*}
Thus, combining the above with eqn.~(\ref{eqn:pd1}), for this single pair of vectors $(x,y) \in \Omega \times \Omega$,
\begin{equation}\label{eqn:eqX33}
\abs{x^TMy} =   \abs{x^TU_A^TDH^TT^T THDU_Ay-x^Ty} \le \frac{1}{2}6\epsilon_1 = 3 \epsilon_1
\end{equation}
holds with probability at least $1-3\delta$. Next, recall that there are no more than $e^{8d}$ pairs of vectors $(x,y)\in\Omega\times\Omega$, and we need eqn.~(\ref{eqn:eqX33}) to hold for all of them. Since we set $\delta=1/(60e^{8d})$ then it follows by a union bound that eqn.~(\ref{eqn:eqX33}) holds for all pairs of vectors $(x,y)\in\Omega\times\Omega$ with probability at least .95. Additionally, let us set $\epsilon_1=3/125$, which implies that $\abs{x^TMy}\leq 9/125 \leq \left(1-2^{-1/2}\right)/4$ thus concluding the proof of the lemma.

Finally, we discuss the values of the parameters $q$ and $k$. Since $\delta = 1/(60e^{8d})$, $\epsilon_1 = 3/125$, and $\alpha = \sqrt{2d \ln(40nd)/n}$, the appropriate values for $q$ and $k$ emerge after elementary manipulations from Lemma~\ref{lem:matousek}.
\end{Proof}

\subsubsection{Satisfying condition~(\ref{eqn:lemma1_ass2})}

In order to prove that condition~(\ref{eqn:lemma1_ass2}) is satisfied, we start with Lemma~\ref{lem:Tmatmult}. In words, this lemma states that given vectors $x$ and $y$ we can use the random sparse projection matrix $T$ to approximate $\abs{x^Ty}$ by $\abs{x^TT^{T}Ty}$, provided that $\VINorm{x}$ (or $\VINorm{y}$, but not necessarily both) is bounded. The proof of this lemma is elementary but tedious and is deferred to Section~\ref{sxn:pf_of_technical_lemma} of the Appendix.
\begin{lemma}
\label{lem:Tmatmult} Let $x,y$ be vectors in $\mathbb{R}^{n}$ such that $\VINorm{x} \le \alpha$. Let $T$ be the $k \times n$ sparse projection matrix of Section~\ref{sxn:review_previous:projecting}, with sparsity parameter $q$. If $q \ge \alpha^2$, then
$$
\Expect{ \abs{x^T T^T Ty-x^Ty}^{2} } \le \frac{2}{k} \VTTNormS{x} \VTTNormS{y} + \frac{1}{k} \VTTNormS{y}.
$$
\end{lemma}
\noindent The following lemma proves that condition~(\ref{eqn:lemma1_ass2}) is satisfied by Algorithm~\ref{alg:alg_projection_fast}. The proof of this lemma depends on the bound provided by Lemma~\ref{lem:Tmatmult}. Recall that $b^{\perp} = U_A^{\perp}{U_A^{\perp}}^{T}b$ and thus $\VTTNorm{b^{\perp}} = \VTTNorm{U_A^{\perp}{U_A^{\perp}}^{T}b} = {\cal Z}$.
\begin{lemma} \label{lem:project_lem40pf}
Assume that eqn.~(\ref{eqn:lem:HU_eqn2}) holds. If $k\geq 60 d/\epsilon$ and $q \geq 2n^{-1} \ln(40nd)$, then, with probability at least .9, $$ \VTTNormS{\left(THDU_A\right)^{T}THDb^{\perp}} \leq \epsilon \mathcal{Z}^{2}/2.$$
\end{lemma}
\begin{Proof}
We first note that since $U_A^Tb^{\perp}=0$, it follows that ${U_A^{(j)}}^{T}b^{\perp}={U_A^{(j)}}^{T}DH^THDb^{\perp}=0$, for all $j\in [d]$. Thus, we have that
\begin{equation}\label{eqn:eqX41}
\VTTNormS{U_A^TDH^T T^T T HDb^{\perp}}
   = \sum_{j=1}^{d} \left( \left(\left(HDU_A\right)^{(j)}\right)^{T}T^TTHDb^{\perp} - {U_A^{(j)}}^{T}DH^THDb^{\perp} \right)^2   .
\end{equation}
We now bound the expectation of the left hand side of eqn.~(\ref{eqn:eqX41}) by using Lemma~\ref{lem:Tmatmult} to bound each term on the right hand side of eqn.~(\ref{eqn:eqX41}). Using eqn.~(\ref{eqn:eqpd12}) of Lemma~\ref{lem:HU} we get that
$$
\VINorm{\left(HDU_A\right)^{(j)}}
   \le   \sqrt{2n^{-1}\ln(40nd)}
$$
holds for all $j\in[d]$. By our choice of the sparsity parameter $q$ the conditions of Lemma~\ref{lem:Tmatmult} are satisfied. It follows from Lemma~\ref{lem:Tmatmult} that
\begin{eqnarray*}
\Expect{ \VTTNormS{U_A^TDH^TT^TTHDb^{\perp} } }  &  =& \sum_{j=1}^{d} \Expect{ \left( \left(\left(HDU_A\right)^{(j)}\right)^{T}T^TTHDb^{\perp} - {U_A^{(j)}}^{T}DH^THDb^{\perp} \right)^2 } \\
&\leq& \sum_{j=1}^d \left(\frac{2}{k}\VTTNormS{\left(HDU_A\right)^{(j)}}\VTTNormS{HDb^{\perp}} + \frac{1}{k}\VTTNormS{HDb^{\perp}}\right)   \\
&=& \frac{3d}{k}\VTTNormS{HDb^{\perp}} = \frac{3d}{k}{\cal Z}^2.
\end{eqnarray*}
The last line follows since $\VTTNorm{\left(HDU_A\right)^{(j)}}=1$, for all $j\in[d]$. Using Markov's inequality, we get that with probability at least $.9$,
$$
\VTTNormS{U_A^TDH^TT^TTHDb^{\perp}} \le\frac{30d}{k}\mathcal{Z}^2.
$$
The proof of the lemma is concluded by using the assumed value of $k$.
\end{Proof}

\subsubsection{Proving Theorem~\ref{thm:alg_projection_fast}}

By our choices of $k$ and $q$ as in eqns.~(\ref{eqn:valuekfinal}) and~(\ref{eqn:valueqfinal}), it follows that both conditions~(\ref{eqn:lemma1_ass1}) and~(\ref{eqn:lemma1_ass2}) are satisfied. Combining with Lemma~\ref{lem:suff_cond} we immediately get the accuracy guarantees of Theorem~\ref{thm:alg_projection_fast}. The failure probability of Algorithm~\ref{alg:alg_projection_fast} can be bounded using an argument similar to the one used in Section~\ref{sxn:sampling:proofs:complete}.

In order to complete the proof we discuss the running time of Algorithm~\ref{alg:alg_projection_fast}. First of all, by the construction of $T$, the expected number of non-zero entries in $T$ is $kqn$. In Step $5$ we need to compute the products $THDA$ and $THDb$. Recall that $A$ has $d$ columns and thus the running time of computing both products is equal to the time needed to apply $THD$ on $(d+1)$ vectors. First, note that in order to apply $D$ on $(d+1)$ vectors in $\mathbb{R}^n$, $n(d+1)$ operations suffice. In order to estimate how many operations are needed to apply $TH$ on $(d+1)$ vectors, we use the results of Theorem~$2.1$ (see also Section 7) of Ailon and Liberty~\cite{AL08}, which state that at most $2n(d+1)\log_2 \left(\abs{T}+1\right)$ operations are needed for this operation. Here $\abs{T}$ denotes the number of non-zero elements in the matrix $T$, which -- in expectation -- is $nkq$. After this preprocessing, Algorithm~\ref{alg:alg_projection_fast} must compute the pseudoinverse of a $k \times d$ matrix, or, equivalently, solve a least-squares problem on $k$ constraints and $d$ variables. This operation can be performed in $O(kd^2)$ time since $k \geq d$. Thus, the entire algorithm runs in expected time
$$n(d+1) + 2n(d+1) \Expect{\log_2 \left(\abs{T} + 1\right)} + O\left(kd^2 \right) \leq n(d+1) + 2n(d+1) \log_2 \left(nkq + 1\right) + O\left(kd^2 \right).$$
%


%
\section{Appendix}

\subsection{Approximating matrix multiplication}

Let $A \in \mathbb{R}^{m \times n}$ be any matrix. Consider the following algorithm (which is essentially the algorithm in page 876 of~\cite{DMM08_CURtheory_JRNL}) that constructs a matrix $C\in \mathbb{R}^{m \times c}$ consisting of $c$ rescaled columns of $A$. We will seek a bound on the approximation error $\TNorm{AA^T-CC^T}$, which we will provide in Theorem~\ref{thm:theorem7correct}. A variant of this theorem appeared as Theorem 7 in~\cite{DMM08_CURtheory_JRNL}; this version modifies and supersedes eqn. (47) of Theorem 7 in the following manner: first, we will assume that the spectral norm of $A$ is bounded and is at most one (this is a minor normalization assumption). Second, and most importantly, we will need to set $c$ to be at least the value of eqn. (\ref{eqn:CboundAppendix}) for the theorem to hold. This second assumption was omitted from the statement of eqn. (47) in Theorem 7 of~\cite{DMM08_CURtheory_JRNL}.

\begin{algorithm}
\begin{framed}

\SetLine

\AlgData{
$A \in \mathbb{R}^{m \times n}$,
$p_i \geq 0, i\in[n]$ s.t. $\sum_{i \in [n]}p_i=1$,
positive integer $c \leq n$.}

\AlgResult{
$C \in \mathbb{R}^{m \times c}$
}

Initialize $S \in \mathbb{R}^{m \times c}$ to be an all-zero matrix.

\For{$t=1,\ldots,c$}{
   Pick $i_t \in [n]$, where $\Prob(i_t = i) = p_i$\;
   $S_{i_t t} = 1/\sqrt{cp_{i_t}}$\;
}

Return $C = AS$\;

\end{framed}
\caption{
The \textsc{Exactly($c$)} algorithm.
}
\label{alg:SDconstruct_exact}
\end{algorithm}

\begin{theorem}\label{thm:theorem7correct}
Let $A \in \mathbb{R}^{m \times n}$ with $\TNorm{A} \leq 1$. Construct $C$ using the \textsc{Exactly($c$)} algorithm and let the sampling probabilities $p_i$ satisfy
\begin{equation}\label{eqn:defPj}
p_i \geq \beta \frac{\TNormS{A^{(i)}}}{\FNormS{A}}
\end{equation}
for all $i \in [n]$ for some constant $\beta \in (0,1]$. Let $\epsilon \in (0,1)$ be an accuracy parameter and assume $\FNormS{A}\geq 1/24$. If
\begin{equation}\label{eqn:CboundAppendix}
c \geq \frac{96 \FNormS{A}}{\beta \epsilon^2}\ln \left(\frac{96 \FNormS{A}}{\beta \epsilon^2 \sqrt{\delta}}\right)
\end{equation}
then, with probability at least $1-\delta$,
$$\TNorm{AA^T-CC^T}\leq \epsilon.$$
\end{theorem}

\begin{Proof}
Consider the \textsc{Exactly}$(c)$ algorithm. Then
\begin{equation*}
AA^T = \sum_{i=1}^n A^{(i)} A^{{(i)}^T}.
\end{equation*}
Similar to~\cite{RV07} we shall view the matrix $AA^T$ as the true mean of a bounded operator valued random variable, whereas $CC^T = AS (AS)^T = ASS^TA^T$ will be its empirical mean. Then, we will apply Lemma 1 of~\cite{Oli10}. To this end, define a random vector $y \in \mathbb{R}^m$ as
\begin{equation*}
\Probab{y = \frac{1}{\sqrt{p_i}}A^{(i)}} = p_i
\end{equation*}
for $i \in [n]$. The matrix $C = AS$ has columns $\frac{1}{\sqrt{c}}y^1,\frac{1}{\sqrt{c}}y^2,\ldots,\frac{1}{\sqrt{c}}y^c$, where $y^1,y^2,\ldots,y^c$ are $c$ independent copies of $y$. Using this notation, it follows that
\begin{equation}\label{eqn:expectyyt}
\Expect{yy^T} = AA^T
\end{equation}
and
\begin{equation*}
CC^T = ASS^TA^T = \frac{1}{c}\sum_{t=1}^c y^t{y^t}^T.
\end{equation*}
Finally, let
\begin{equation}\label{eqn:defM}
M = \TNorm{y} = \frac{1}{\sqrt{p_i}}\TNorm{A^{(i)}}.
\end{equation}
We can now apply Lemma 1, p. 3 of~\cite{Oli10}. Notice that from eqn. (\ref{eqn:expectyyt}) and our assumption on the spectral norm of $A$, we immediately get that $$\TNorm{\Expect{yy^T}} = \TNorm{AA^T} \leq \TNorm{A}\TNorm{A^T} \leq 1.$$ Then, Lemma 1 of~\cite{Oli10} implies that
\begin{equation}\label{eqn:ExpectBound}
\TNorm{CC^T-AA^T} < \epsilon,
\end{equation}
with probability at least $1-\left(2c\right)^2 \exp\left(-\frac{c\epsilon^2}{16M^2 + 8M^2 \epsilon}\right)$. Let $\delta$ be the failure probability of Theorem~\ref{thm:theorem7correct}; we seek an appropriate value of $c$ in order to guarantee $\left(2c\right)^2 \exp\left(-\frac{c\epsilon^2}{16M^2 + 8M^2 \epsilon}\right) \leq \delta$. Equivalently, we need to satisfy
$$\frac{c}{\ln \left(2c/\sqrt{\delta}\right)} \geq \frac{2}{\epsilon^2}\left(16M^2 + 8M^2\epsilon\right).$$
Recall that $\epsilon < 1$, and combine eqns. (\ref{eqn:defM}) and (\ref{eqn:defPj}) to get $M^2 \leq \FNormS{A}/\beta$. Combining with the above equation, it suffices to choose a value of $c$ such that
$$\frac{c}{\ln \left(2c/\sqrt{\delta}\right)} \geq \frac{48}{\beta\epsilon^2}\FNormS{A},$$
or, equivalently,
$$\frac{2c/\sqrt{\delta}}{\ln \left(2c/\sqrt{\delta}\right)} \geq \frac{96}{\beta\epsilon^2\sqrt{\delta}}\FNormS{A}.$$
We now use the fact that for any $\eta \geq 4$, if $x \geq 2\eta \ln \eta$ then $\frac{x}{\ln x} \geq \eta$. Let $x = 2c/\sqrt{\delta}$, let $\eta = 96 \FNormS{A}/\left(\beta \epsilon^2\sqrt{\delta}\right)$, and note that $\eta \geq 4$ if $\FNormS{A} \geq 1/24$, since $\beta$, $\epsilon$, and $\delta$ are at most one. Thus, it suffices to set
$$\frac{2c}{\sqrt{\delta}} \geq 2 \frac{96 \FNormS{A}}{\beta \epsilon^2\sqrt{\delta}}\ln \left(\frac{96 \FNormS{A}}{\beta \epsilon^2\sqrt{\delta}}\right),$$
which concludes the proof of the theorem.
\end{Proof}

\subsection{The proof of Lemma~\ref{lem:Tmatmult}}\label{sxn:pf_of_technical_lemma}

Let $T \in \mathbb{R}^{k \times n}$ be the sparse projection matrix constructed via Algorithm~\ref{alg:alg_projection_fast} (see Section~\ref{sxn:projection:result}), with sparsity parameter $q$. In addition, given $x,y\in\mathbb{R}^{n}$, let $\Delta = x^TT^TTy-x^Ty$. We will derive a bound for $$\Expect{\Delta^2} = \Expect{\left(x^TT^TTy-x^Ty\right)^2}.$$
Let $t_{(i)}$ be the $i$-th row of $T$ \textit{as a row vector}, for $i\in[k]$, in which case
$$
\Delta = \sum_{i=1}^{k} \left( x^Tt_{(i)}^Tt_{(i)}y-\frac{1}{k}x^Ty \right)  .
$$
Rather than computing $\Expect{\Delta^2}$ directly, we will instead use that $\Expect{\Delta^2}=\left(\Expect{\Delta}\right)^2+\Varnce{\Delta}$. We first claim that $\Expect{\Delta}=0$. By linearity of expectation,
\begin{equation}
\label{eqn:technical_pr_eq10}
\Expect{\Delta}
   = \sum_{i=1}^{k} \left[ \Expect{x^Tt_{(i)}^Tt_{(i)}y}-\frac{1}{k}x^Ty  \right]   .
\end{equation}
We first analyze
$t_{(i)}=t$ for some fixed $i$ (w.l.o.g. $i=1$). Let $t_i$ denote the $i$-th element of the vector $t$ and recall that $\Expect{t_i}=0$, $\Expect{t_i t_j}=0$ for $i \ne j$,
and also that $\Expect{t_i^2}=1/k$.
Thus,
\begin{equation*}
\Expect{x^Tt^Tty}
   = \Expect{\sum_{i=1}^{n}\sum_{j=1}^{n} x_it_it_jy_j}
   = \sum_{i=1}^{n}\sum_{j=1}^{n} x_i \Expect{t_it_j} y_j
   = \sum_{i=1}^{n} x_i \Expect{t_i^2} y_i
   = \frac{1}{k} x^Ty   .
\end{equation*}
By combining the above with eqn.~(\ref{eqn:technical_pr_eq10}), it follows that
$\Expect{\Delta}=0$, and thus that $\Expect{\Delta^2}=\Varnce{\Delta}$.
In order to provide a bound for $\Varnce{\Delta}$, note that
\begin{eqnarray}
\label{eqn:technical_pr_eq20}
\Varnce{\Delta} &=& \sum_{i=1}^{k} \Varnce{ x^Tt_{(i)}^Tt_{(i)}y - \frac{1}{k}x^Ty } \\
\label{eqn:technical_pr_eq30}
            &=& \sum_{i=1}^{k} \Varnce{ x^Tt_{(i)}^Tt_{(i)}y }.
\end{eqnarray}
Eqn.~(\ref{eqn:technical_pr_eq20}) follows since the $k$ random variables
$x^Tt_{(i)}^Tt_{(i)}y-\frac{1}{k}x^Ty$ are independent (since the elements
of $T$ are independent) and eqn.~(\ref{eqn:technical_pr_eq30})
follows since $\frac{1}{k}x^Ty$ is constant.
In order to bound eqn.~(\ref{eqn:technical_pr_eq30}), we first analyze
$t_{(i)}=t$ for some $i$ (w.l.o.g. $i=1$).
Then,
\begin{eqnarray}
\nonumber
\Varnce{x^Tt^Tty}
   &=& \Expect{(x^Tt^Tty)^2} - \left(\Expect{x^Tt^Tty}\right)^2    \\
\label{eqn:eqX51}
   &=& \Expect{(x^Tt^Tty)^2} - \frac{1}{k^2}(x^Ty)^2      .
\end{eqnarray}
We will bound the $\Expect{(x^Tt^Tty)^2}$ term directly:
\begin{eqnarray}
\nonumber
\Expect{\left(\sum_{i=1}^{n}\sum_{j=1}^{n}x_it_it_jy_j\right)^2}
   &=& \Expect{\sum_{i_1=1}^{n}\sum_{i_2=1}^{n}\sum_{j_1=1}^{n}\sum_{j_2=1}^{n}x_{i_1}x_{i_2}t_{i_1}t_{i_2}t_{j_1}t_{j_2}y_{j_1}y_{j_2} } \\
\label{eqn:star1}
   &=& \sum_{i_1=1}^{n}\sum_{i_2=1}^{n}\sum_{j_1=1}^{n}\sum_{j_2=1}^{n}x_{i_1}x_{i_2} \Expect{ t_{i_1}t_{i_2}t_{j_1}t_{j_2} } y_{j_1}y_{j_2}    .
\end{eqnarray}
Notice that if any of the four indices $i_1,i_2,j_1,j_2$ appears only once,
then the expectation $\Expect{ t_{i_1}t_{i_2}t_{j_1}t_{j_2} }$ corresponding
to those indices equals zero.
This expectation is non-zero if the four indices are paired in couples or
if all four are equal.
That is, non-zero expectation happens if
\begin{eqnarray*}
\mbox{(A)} &:& i_1=i_2 \neq j_1=j_2 \hspace{1.0cm} \mbox{ ($n^2-n$ terms)}   \\
\mbox{(B)} &:& i_1=j_1 \neq i_2=j_2 \hspace{1.0cm} \mbox{ ($n^2-n$ terms)}   \\
\mbox{(C)} &:& i_1=j_2 \neq i_2=j_1 \hspace{1.0cm} \mbox{ ($n^2-n$ terms)}   \\
\mbox{(D)} &:& i_1=i_2 =    j_1=j_2 \hspace{1.0cm} \mbox{ ($n$ terms)}   .
\end{eqnarray*}
For case~(A), let $i_1=i_2=\ell$ and let $j_1=j_2=p$, in which case
the corresponding terms in eqn.~(\ref{eqn:star1}) become:
\begin{eqnarray*}
\sum_{\ell=1}^{n}\sum_{p=1:p\ne\ell}^{n} x_{\ell}^2\Expect{t_{\ell}^2t_p^2}y_p^2
   &=& \sum_{\ell=1}^{n}\sum_{p=1:p\ne\ell}^{n} x_{\ell}^2\Expect{t_{\ell}^2}\Expect{t_p^2}y_p^2   \\
   &=& \frac{1}{k^2} \sum_{\ell=1}^{n}\sum_{p=1:p\ne\ell}^{n} x_{\ell}^2y_p^2   \\
   &=& \frac{1}{k^2} \sum_{\ell=1}^{n}\sum_{p=1:p\ne\ell}^{n} x_{\ell}^2y_p^2 + \frac{1}{k^2} \sum_{p=1}^{n} x_p^2y_p^2 - \frac{1}{k^2} \sum_{p=1}^{n}x_p^2y_p^2    \\
   &=& \frac{1}{k^2} \VTTNormS{x} \VTTNormS{y} - \frac{1}{k^2} \sum_{p=1}^n x_p^2 y_p^2    .
\end{eqnarray*}
Similarly, cases~(B) and~(C) give:
\begin{eqnarray*}
\sum_{\ell=1}^{n}\sum_{p=1:p\ne\ell}^{n} x_{\ell}x_p\Expect{t_{\ell}^2t_p^2}y_{\ell}y_p
   &=& \frac{1}{k^2}(x^Ty)^2 - \frac{1}{k^2} \sum_{p=1}^{n} x_p^2 y_p^2   \\
   & & \mbox{(where $i_1 = j_1 = \ell$ and $i_2 = j_2 = p$)}  , \mbox{ and} \\
\sum_{\ell=1}^{n}\sum_{p=1:p\ne\ell}^{n} x_{\ell}x_p\Expect{t_{\ell}^2t_p^2}y_{\ell}y_p
   &=& \frac{1}{k^2}(x^Ty)^2 - \frac{1}{k^2} \sum_{p=1}^{n} x_p^2 y_p^2   \\
   & & \mbox{(where $i_1 = j_2 = \ell$ and $i_2 = j_1 = p$)}          .
\end{eqnarray*}
Finally, for case~(D), let $i_1=i_2=j_1=j_2=\ell$, in which case:
$$
\sum_{\ell=1}^{n} x_{\ell}^2\Expect{t_{\ell}^4}y_{\ell}^2
   = \frac{1}{k^2q} \sum_{\ell=1}^{n} x_{\ell}^2y_{\ell}^2   ,
$$
where we have used that $\Expect{t_{\ell}^4}=1/(k^2q)$.
By combining these four terms for each of the $k$ terms in the sum, it follows
from eqns.~(\ref{eqn:technical_pr_eq30}) and~(\ref{eqn:eqX51}) that
\begin{eqnarray}
\nonumber
\Expect{\Delta^2}
   &=& k\left( \frac{1}{k^2}\VTTNormS{x}\VTTNormS{y}
             + \frac{2}{k^2}(x^Ty)^2
             - \frac{3}{k^2} \sum_{p=1}^n x_p^2 y_p^2
             + \frac{1}{k^2q}\sum_{p=1}^n x_p^2 y_p^2
             - \frac{1}{k^2}(x^Ty)^2
        \right)     \\
\label{eqn:eqX61}
   &\leq& \frac{2}{k}\VTTNormS{x}\VTTNormS{y}
        + \frac{1}{kq} \sum_{p=1}^n x_p^2 y_p^2.
\end{eqnarray}
In the above we used $(x^Ty)^2 \le \VTTNormS{x}\VTTNormS{y}$.
Since we assumed that $\VINorm{x}\le\alpha$, the second term on the right
hand side of eqn.~(\ref{eqn:eqX61}) is bounded by
$\frac{\alpha^2}{kq}\VTTNormS{y}$ and the lemma follows since we have
assumed that $q \ge \alpha^2$.
\end{document}